\documentclass[
superscriptaddress,
reprint, aps, prl,
floatfix
]{revtex4-2}

\usepackage{graphicx}
\usepackage{dcolumn}
\usepackage{bm}

\usepackage[utf8]{inputenc}
\usepackage[T1]{fontenc}
\usepackage{mathptmx}
\DeclareMathAlphabet{\mathcal}{OMS}{cmsy}{m}{n}

\usepackage{braket}
\usepackage{amsmath}
\usepackage{caption}
\usepackage{subcaption}
\usepackage{xcolor}
\usepackage{graphicx}
\usepackage{mwe}
\usepackage{stackengine}

\usepackage{physics}

\usepackage{setspace}

\newcommand{\uncertainty}[2]{\raisebox{0.5ex}{\tiny$^{+#1}_{-#2}$}}

\newcommand{\trapDepthNoMHz}{8.9(2)} 
\newcommand{\trapDepth}{\trapDepthNoMHz\,MHz}

\newcommand{\fidelity}{99.91\uncertainty{0.02}{0.02}\%}
\newcommand{\infidelity}{0.09\uncertainty{0.02}{0.02}\%}
\newcommand{\brightError}{0.11\uncertainty{0.03}{0.02}\%} 
\newcommand{\darkError}{0.06\uncertainty{0.01}{0.02}\%} 
\newcommand{\brightErrorN}{107}
\newcommand{\darkErrorN}{64}
\newcommand{\loss}{0.9(2)\%}

\newcommand{\fidelityFibers}{99.84\uncertainty{0.02}{0.03}\%}

\newcommand{\collectionEfficiencyFibers}{0.37\%}

\newcommand{\lossFibers}{2.6(2)\%}

\newcommand{\tOne}{0.49(3)\,s} 
\newcommand{\tOneDark}{0.58(3)\,s} 
\newcommand{\tOneBright}{0.41(3)\,s} 
\newcommand{\highLossFidelity}{99.89\uncertainty{0.02}{0.02}\%}

\newcommand{\highLossLoss}{14.1(3)\%}

\newcommand{\freeSpaceBrightError}{0.22\uncertainty{0.03}{0.03}\%} 
\newcommand{\freeSpaceDarkError}{0.13\uncertainty{0.02}{0.02}\%} 
\newcommand{\freeSpaceInfidelity}{0.18\uncertainty{0.02}{0.02}\%}
\newcommand{\freeSpaceLoss}{1.8(3)\%}

\newcommand{\coolTemperature}{10\,$\mu$K}

\newcommand{\collectionEfficiency}{0.96(1)\%} 
\newcommand{\sigmaDepumpProbability}{$50(5)\times10^{-6}$}
\newcommand{\piDepumpProbability}{$7(4)\times10^{-6}$}

\newcommand{\darkRate}{5.2(3)$\times10^{2}$\,s$^{-1}$} 
\newcommand{\brightRate}{5.44(3)$\times10^{4}$\,s$^{-1}$} 
\newcommand{\idealDarkError}{0.04\%} 
\newcommand{\idealBrightError}{0.005\%} 
\newcommand{\idealTime}{0.26\,ms}

\newcommand{\probeOffResRate}{2.3\,s$^{-1}$}
\newcommand{\probeOffResProbability}{$5.1\times 10^{-5}$}

\newcommand{\D}[1]{$D_{#1}$} 

\begin{document}
	
\preprint{AIP/123-QED}

\title[]{High-Fidelity, Low-Loss State Detection of Alkali-Metal Atoms in Optical Tweezer Traps}

\author{Matthew N. H. Chow}
\email{mnchow@sandia.gov}
\affiliation{Sandia National Laboratories, Albuquerque, New Mexico 87185, USA}
\affiliation{Department of Physics and Astronomy, University of New Mexico, Albuquerque, New Mexico 87106, USA}
\affiliation{Center for Quantum Information and Control, University of New Mexico, Albuquerque, New Mexico 87131, USA}
\author{Bethany ~J. Little}
\affiliation{Sandia National Laboratories, Albuquerque, New Mexico 87185, USA}
\author{Yuan-Yu Jau}
\affiliation{Sandia National Laboratories, Albuquerque, New Mexico 87185, USA}
\affiliation{Department of Physics and Astronomy, University of New Mexico, Albuquerque, New Mexico 87106, USA}
\affiliation{Center for Quantum Information and Control, University of New Mexico, Albuquerque, New Mexico 87131, USA}

\date{\today}

\begin{abstract}
    We demonstrate discrimination of ground-state hyperfine manifolds of a cesium atom in an optical tweezer using a simple probe beam with \fidelity{} detection fidelity and \loss{} detection-driven loss of bright state atoms. Our detection infidelity of \infidelity{} is an order of magnitude better than previously published low-loss readout results for alkali-metal atoms in optical tweezers. Our low atom loss and high-fidelity state detection eliminates the extra depumping mechanism due to population transfer between excited-state sublevels through V-type stimulated Raman transitions caused by the trap laser when the probe laser is present. In this work, complex optical systems and stringent vacuum pressures are not required.
\end{abstract}

\maketitle

\begin{figure}
     \centering
     \includegraphics[trim=0cm 0.4cm 0.5cm 1.2cm, width=0.45\textwidth]{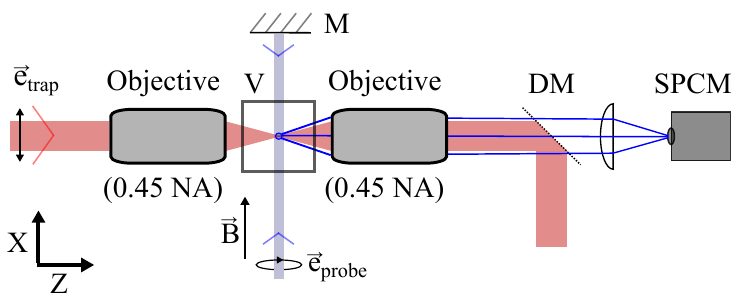}
    \caption{ We use a simple probe beam and fluorescence collection optical system to measure high-fidelity state detection. Both the probe beam (blue shaded line) propagation direction and trap light (red shaded line) polarization ($\vec{\text{e}}_{\text{trap}}$) are oriented along the bias magnetic field direction ($\vec{\text{B}}$). The probe is right-hand circularly polarized in order to drive only $\sigma_{+}$ transitions. The probe is retroreflected with a gold mirror (M) to minimize polarization distortions. Fluorescence (blue rays) is collected outside of the vacuum chamber (V) with a  0.45\,NA microscope objective, separated from the trap light with a dichroic mirror (DM), and focused onto a single-photon counting module (SPCM).
    }
    \label{fig:setup}
\end{figure}

Neutral atoms in arrays of optical dipole traps (ODTs) provide a promising platform for quantum computing \cite{EndresHighFidelity, LukinHighFidelity, LukinCoherentTransportQC}, quantum simulation \cite{Lukin51AtomSimulator, BrowaeysRydbergIsing}, quantum chemistry \cite{NiSingleMolecule}, and optical clocks \cite{TweezerClock}. Atoms can be individually controlled inside tightly focused beams, known as optical tweezers, and have been used to generate defect-free arrays of tens to hundreds of atoms in one-, two- and three-dimensional geometries \cite{EndresDefectFree, BrowaeysDefectFree, Birkl100AtomArray, Browaeys3DArray}. Alkali-metal atoms are frequently utilized because of the simplicity of their electronic structure and well-established methods of laser cooling \footnote{Alkaline earth and Helium-like atoms have become an increasingly popular choice as well, with the most recent progress in Strontium and Ytterbium, which have attractive, narrow line optical clock transitions.}. 

Performance of alkali-metal atoms in tweezer platforms has to date been limited by state readout fidelity.  Detection schemes are typically based on either state-dependent fluorescence collection or loss, where the state of the atom is mapped to trap occupation.
Although relatively high detection fidelity has been achieved with loss-based schemes \cite{LukinCoherentTransportQC}, these methods impose a vacuum-dependent upper bound on readout fidelity, slow the repetition rate, and complicate algorithms requiring mid-circuit measurement.  Fluorescence detection on the other hand allows -- in principle -- for high fidelity state measurement without losing the atom.  Previously, results for alkali atoms with fluorescence detection have shown no better than 1.2\% infidelity without enhancement from an optical cavity \cite{BrowaeysStateDet, SaffmanStateDet, MeschedeBayesianStateDet, MoehringCavity, ReichelCavity, DanCavity}. 
Alternately, high-fidelity, low-loss detection of alkali atoms has been demonstrated in an optical lattice using a state-dependent potential method \cite{WeissSternGerlach}, but adaptation of this method to optical tweezers is not straightforward.

In this paper, we report state discrimination between ground-state hyperfine manifolds of Cs atoms with \infidelity{} infidelity while suffering only \loss{} detection-driven losses of the bright state. 
We achieve these results by implementing an adaptive detection scheme to manage detection heating and by mitigating a previously unreported depumping channel stemming from simultaneous probe and trap illumination to minimize state information loss. We do not impose any stringent vacuum or optical system requirements to achieve this result, enabling straightforward integration of our techniques on contemporary alkali-metal atom tweezer platforms. 

This work is an important step towards scalable, high performance alkali-tweezer machines. While the measurement error on noisy intermediate-scale quantum (NISQ) computers is often overlooked as it does not typically scale with algorithm length, it \textit{is} important to consider the  (generally exponential) scaling of the measurement error with system size.
In addition, by demonstrating simultaneous high-fidelity and low loss, we improve the outlook for quantum sensing applications where the sensitivity is tied to atom retention capability via repetition rate.

\begin{figure*}
     \centering
     \includegraphics[width=\textwidth]{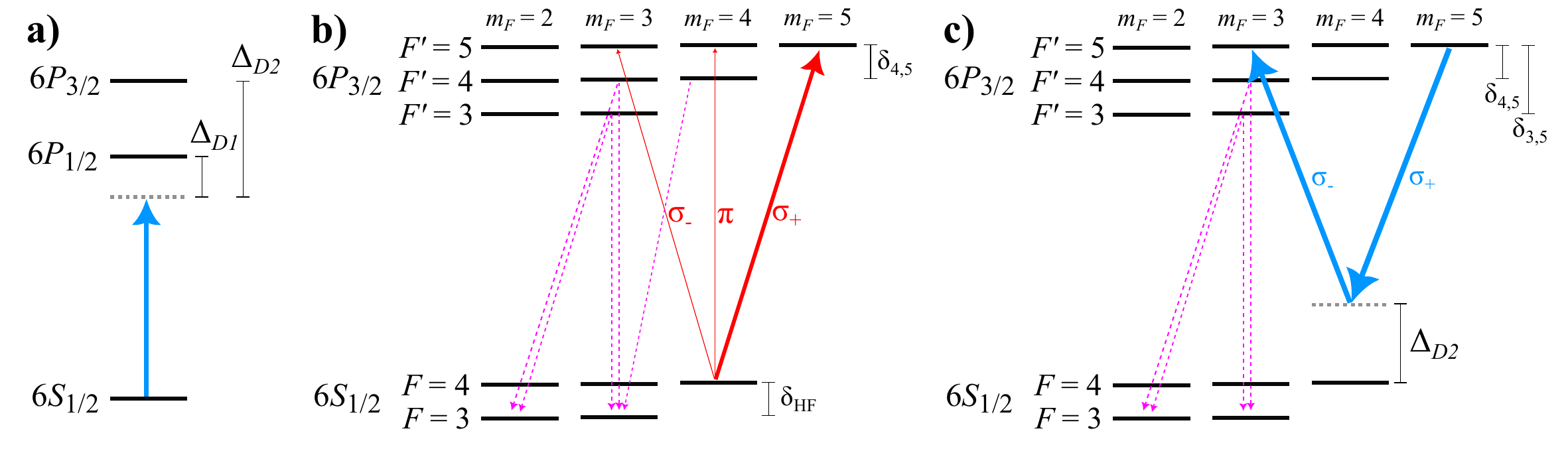}
    \caption{Three mechanisms contribute to state information loss  during detection. \textbf{(a)} The fundamental limitation of effective $T_{1}$ relaxation in an optical dipole trap is imposed by the off-resonant scattering rate from the light used to generate the trapping potential (solid blue line). For trapping parameters used in this manuscript ($U_{0}/h$ = \trapDepth{}), $T_{1}$ is measured to be \tOne{}. \textbf{(b)} Probe (solid red lines) polarization impurities and misalignment to the magnetic field axis allow for weak, off-resonant coupling to $F'\neq$5 excited state hyperfine manifolds, which then have dipole allowed decay channels (dashed lines) to the \textit{F}=3 ground state manifold. \textbf{(c)}  $\sigma_{\pm}$ trap light (solid blue lines) allows for detuned, two-photon coupling from $F'$=5 to $F'\neq$5 excited state hyperfine manifolds. This depumping mechanism can be effectively eliminated by using $\pi$-polarized trap light, for which coupling between $F'=5, m_{F}'=5$ and the ground state is not supported. Levels not drawn to scale.
    }
    \label{fig:det_lvl}
\end{figure*}

Our experimental sequence is triggered upon loading a Cs atom from a magneto-optical trap into a tweezer trap. The tweezer is formed by focusing up to 10\,mW of linearly polarized 937\,nm laser light through a 0.45\,NA microscope objective (OptoSigma PAL-20-NIR-LC00) to a spot with a $1.6\,\mu$m$ \, 1/e^{2}$ waist radius. The trap is small enough that a light-assisted collisional-blockade mechanism ensures loading no more than one atom at a time \cite{AndersenSingleAtom}. The trap wavelength of $\lambda_{\rm trap}=$ 937\,nm is chosen to be red-detuned from both the \D{1} and \D{2} line transitions and to have approximately the same AC Stark shift on both states (6$S_{1/2}$ and 6$P_{3/2}$) of the cooling transition (\textit{i.e.} ``magic''). 
After loading an atom, we cool it with polarization-gradient cooling to $\approx$ \coolTemperature{}. The atom is then prepared in either $\ket{F=4}$ (``bright'') or $\ket{F=3}$ (``dark'') hyperfine ground manifolds and the state of the atom is then read out using a near-resonant probe laser. 
Finally, we check for atom retention in the trap with the cooling beams.

During detection, we use a near resonant probe beam that is tuned to 
the \textit{F}=4 to \textit{F}$'$=5 \D{2} line transition so the \textit{F}=4, bright ground state scatters probe photons in a closed cycling transition, while the \textit{F}=3 state is dark to the probe and ideally scatters no light \footnote{Intensity of the probe is 0.40(1)$I_{\rm sat}$ where $I_{\rm sat}$ is the saturation intensity on the $\ket{F=4,m_F=5}\rightarrow\ket{F'=5,m'_{F}=5}$ transition. Detuning of the probe is 0.26(3) MHz red of the same line.}. State-dependent fluorescence is then collected and imaged onto a single photon counting module (SPCM, Excelitas SPCM-AQRH-16), which is monitored in real time by our field programmable gate-array (FPGA) control system. The atom is assigned a label, bright or dark, based on a discrimination threshold level of collected photon counts. The probe beam is right-hand circularly polarized and a bias magnetic field (B-field) of 4\,G is set along its propagation direction such that we drive primarily $\sigma_{+}$ transitions. Thus, we quickly pump the atom into the $F=4, \ m_{F}=4$ ``stretched" ground state, which is protected via angular momentum conservation against depumping to the dark state through off-resonant coupling to $F'\neq5$ states.
There is also a small, transient probability of depumping during the driven random walk up to the stretched state if the atom is initialized in a different Zeeman state - see Ref. \cite{SaffmanStateDet}. For this reason, we turn on the probe beam alongside the repump beam in the last 10\,$\mu s$ of optical pumping when preparing the bright state, effectively preparing the stretched state \footnote{We also retroreflect the probe beam to balance the recoil force and use a gold mirror to minimize polarization distortion.}.

In an ideal scenario, we would be able to determine the state of the atom with arbitrary accuracy by collecting scattered photons until the bright and dark count distributions are sufficiently separated. However, recoil heating of the bright state during detection limits the maximum possible probe time before atom loss, and depumping from trap and probe light can lead to state information loss during detection. We study three depumping mechanisms, illustrated in Fig.~\ref{fig:det_lvl}: (a) off-resonant scatter of trap light, (b) probe polarization components that allow for off-resonant scatter via $F'\neq5$, and (c) trap-induced two-photon coupling of excited state hyperfine manifolds via stimulated Raman transitions. 
The first mechanism is the main fundamental constraint, and can be mitigated by improving the ratio of photon collection to trap off-resonant scattering rates. The second mechanism is imposed by probe misalignment and polarization impurity, and can be improved with further technical capability. The third is an almost entirely geometrical problem and can be essentially eliminated by an appropriate choice of trap polarization, bias fields, probe polarization, and probe propagation direction.  To our knowledge, this third information loss channel has not been previously reported for this type of detection, and mitigation of this depumping pathway turned out to be critical to achieving the state discrimination fidelity observed in our work.

To guide our choice of detection parameters, we use a statistical model of photon collection governed by the rates of three processes: the count collection rate from a bright atom, the dark collection rate (background), and the rate of state-information loss from the prepared state of the atom (the depump rate). 
We use a short detection time compared to the depump rate and thus consider at most a single state change event during detection \footnote{We initially considered the possibility of multiple state change events, however that is suppressed to first order in $t_{d}R_{dep}$.}. The resulting photon collection distribution is found by marginalizing over state change event times, $t$:
\begin{equation}
    \label{eq:det-model}
    \begin{split}
    P(n) = \ &e^{-t_{d}R_{\rm dep}}\mathcal{P}(n, R_{p}t_{d}) \\
    &+ \int_{0}^{t_{d}} \mathcal{P}(n, R_{p}t + R_{np}(t_{d}-t)) R_{\rm dep}e^{-tR_{\rm dep}} dt.
    \end{split}
\end{equation}

Here, $P(n)$ is the probability of detecting $n$ photons, $\mathcal{P}(n, Rt) = e^{-Rt}(Rt)^{n}/n!$ is probability of $n$ events in a Poisson distribution of mean $Rt$, $t_{d}$ is the total detection time, $R_{p}$ is the photon collection rate from an atom in the prepared state, $R_{np}$ is the collection rate from the atom after a state changing event (i.e., collection rate from the non-prepared state),
and $R_{\rm dep}$ is the depumping rate out of the prepared state.
The first term represents the contribution from cases where no depumping occurs, and the second term is the convolution of counts from before and after a state change event that occurs during the detection window. 

Thresholding at $m$ detected counts to assign state labels, we define the bright label probability as a function of $F$, the prepared state: $P_{\rm bright}(F) = \sum_{n=m}^{\infty}P(n|F)$. The bright state readout error is the probability of failing to assign a bright label to an atom prepared in the bright state: $\epsilon_{\rm bright} = 1 - P_{\rm bright}(F=4)$, and similarly the dark state readout error is the probability of assigning a bright label to an atom prepared in the dark state: $\epsilon_{\rm dark} = P_{\rm bright}(F=3)$.
For all infidelities quoted in this work, we use $\mathcal{I} = \frac{1}{2}(\epsilon_{\rm bright}+\epsilon_{\rm dark})$, and all fidelities are $\mathcal{F}=1-\mathcal{I}$. Numerically, we find that the optimal threshold level is $m=3$ counts to discriminate between bright and dark in the range of collection and depumping rates near our experiment parameters. 

The background photon collection rate is \darkRate{} and the bright atom collection rate is \brightRate{}, such that optimal error rates of $\epsilon_{\rm bright}=$\,\idealBrightError{} and $\epsilon_{\rm dark}=$\,\idealDarkError{} would be achieved in an ideal, depump-free, case at \idealTime{} of probe time. However, observed error rates are significantly higher (especially for the bright state), indicating that state information loss during detection is the dominant source of detection infidelity. As a figure of merit, we consider the rate of information loss, $ R_{\rm dep}$, compared to the rate of information gain, $R_{\rm bright}$: $\mathcal{R} = R_{\rm dep}/R_{\rm bright}$. $\mathcal{R}$ is the depump probability per collection event from the bright state.
As a rough approximation for estimating fidelity, one can consider the probability of collecting $m$ photons from the bright state prior to depump, $(1-\mathcal{R})^{m} \approx 1-m\mathcal{R}$ for $m\mathcal{R} \ll 1$.

We begin our analysis of state information loss during detection by considering single photon off-resonant scattering from the trapping laser, illustrated in Fig.~\ref{fig:det_lvl}a. 
Using a numeric density matrix model including all magnetic levels of the ground and first excited states ($6P_{1/2}$ \& $6P_{3/2}$), we find that the depumping rate from the trap light is $R_{\rm dep, trap} \approx 2.5 \times 10^{-7} U_{0}/h$ \cite{JauBook}. For a trap depth of $U_{0}/h=$\trapDepth{}, this yields an effective $T_{1} = 1/R_{\rm dep, trap}$ of 0.45\,s and a normalized depump rate of $\mathcal{R}_{\rm trap} = 4.1 \times 10^{-5}$. To experimentally verify, we measure exponential decay times for the bright (dark) state and find \tOneBright{} (\tOneDark{}) \footnote{Note that in the case that the atom temperature is significant compared to the trap depth, atomic wavepacket broadening should also be considered when calculating the trap off-resonant scattering rate. See Supplemental Material for details.}.

The probe beam may also cause state information loss due to off-resonant scatter, as illustrated in Fig.~\ref{fig:det_lvl}b.  
Although we attempt to drive only $\sigma_{+}$ transitions with the probe beam, polarization impurity or misalignment of the probe propagation direction to the B-field allows for off-resonant scatter via $F'\neq 5$ levels. Following the calculation in Ref. \cite{SaffmanStateDet}, we find the depump rate by summing over all dipole allowed transition scattering rates, weighted by the branching ratio ($b_{\alpha}$) of the excited state ($\ket{\alpha}$) to the \textit{F}=3 ground state manifold \cite{SupplementalMaterial}. 
The sum depends on probe polarization purity, alignment to the B-field, and intensity. Our measured degree of probe polarization is 99.4\% out of the fiber launch, and the probe reflects off of a single, unprotected gold mirror before entering the vacuum chamber. We align the B-field by scanning shim fields and maximizing the bright label probability for $F=4$ after a conservatively long detection time, such that we expect an alignment tolerance of a few degrees and a resulting 
$\sigma_{-}$- and $\pi$-polarization intensity fractions of $\leq$ 2\%. This corresponds to a probe off-resonant scatter rate of $\leq$ \probeOffResRate{}, or $\mathcal{R}_{\rm probe} \leq $ \probeOffResProbability{}; comparable to the depump probability due to the trap off resonant scatter.

Initially our measured depump rate was higher than we would expect from trap and probe off-resonant scatter mechanisms, which prompted investigation of a third mechanism. A trap-light driven, detuned, two-photon effect, shown in Fig.~\ref{fig:det_lvl}c, provides another pathway for atoms to escape the cycling transition subspace. While the ground state hyperfine splitting of $\approx$ 9.2\,GHz is sufficiently large to prevent significant $\Lambda$-type Raman transitions, the excited state hyperfine splitting is only of order 100\,MHz. For some choices of experiment geometry, this leads to appreciable population leakage of exited state atoms to $F'\neq 5$ by detuned V-type Raman transitions.

As a case study for this V-type Raman mechanism, we calculate the two-photon depumping rate when the bias B-field is aligned orthogonal to the (linear) trap polarization direction, producing ($\sigma_{+}$ + $\sigma_{-}$)/$\sqrt{2}$ trap light in the atom's angular momentum basis. Using three-level calculations and a far-detuned approximation \cite{SupplementalMaterial}, we estimate the effective two-photon Rabi rate for the Raman transition from $\ket{F'=5, m_{F}'=5}$ to $\ket{F'=4(3), m_{F}'=3}$ to be ${\Omega_{{\rm eff},4'}=2\pi \times [-0.255\,U_{0}/h]}$ rad/s ($\Omega_{{\rm eff},3'}=2\pi \times 0.329\,U_{0}/h$ rad/s). Considering the branching ratio of each state to the $\ket{F=3}$ ground state, for our trap depth this gives a depump probability per resonant scattering event of $6.6\times10^{-5}$, or a normalized depump rate of $\mathcal{R}_{\rm Raman} = 6.8 \times 10^{-3}$ after accounting for our measured collection efficiency (CE) of \collectionEfficiency{} \footnote{Collection efficiency quoted here includes transmission through all optics and the quantum efficiency of the detector.}. Since this value is more than an order of magnitude larger than both the probe and trap single photon off-resonant depumping rates, mitigation of this mechanism yields significant gains for the bright state readout fidelity. 

\begin{figure}
    \includegraphics[trim=1.7cm 0.5cm 0.8cm -0.15cm, width=0.4\textwidth]{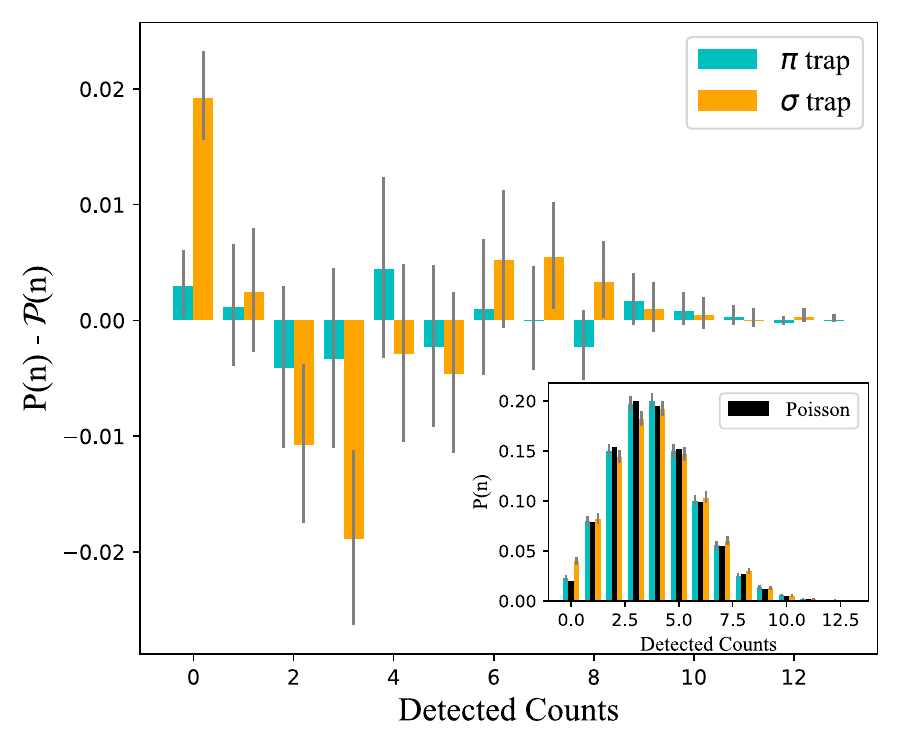}
    \caption{Bright state detection count histograms for different experiment geometries demonstrate the effect of two-photon depumping from the trap light. In the main figure, we take the difference of detection count histograms for a fixed probe duration, $P(n)$, and the depump-free case of a Poisson distribution with the same mean, $\mathcal{P}(n)$. Full histograms are shown in the inset. The distribution collected using the $\sigma$-polarized configuration (orange bars) is significantly non-Poissonian and has a substantially larger zero bin than that of the $\pi$-polarized configuration (teal bars). This indicates a higher depump rate in the $\pi$-polarized configuration that we attribute to V-type Raman transitions from the trap light. When the trap has $\sigma_{\pm}$ components, two-photon coupling between excited manifold hyperfine states through the $F=4, m_{F}=4$ ground state allows for leakage out of the cycling transition subspace, while in the $\pi$-polarized trap configuration, the lack of dipole allowed coupling from the $F'=5, m_{F}'=5$ excited state to $F=4, m_{F}=4$ ground state prevents two-photon leakage during bright state detection. There are 10,000 shots in each data histogram; uncertainty markers are Wilson score intervals. 
    }
    \label{fig:sigma_pi_histograms}
\end{figure}

To verify this newly identified mechanism, we study the bright state count histogram for two experiment geometries. We orient the B-field and probe jointly defined quantization axis either parallel or orthogonal to the trap polarization generating either $\pi$ or $\sigma_{\pm}$ polarization components of the trapping beam \footnote{Neglecting polarization distortion due to focusing and incorrect polarization components due to misalignment.}. With the $\sigma$-polarized trap light, we expect to see depumping from the two-photon pathway as calculated. However, for the $\pi$-polarized trap configuration, there is no dipole-allowed coupling from $\ket{F'=5, m'_F=5}$ to the ground state caused by the trap light, and thus we expect this mechanism to be absent (barring trap polarization misalignment and polarization distortion effects from focusing). 

As shown in Fig.~\ref{fig:sigma_pi_histograms}, we run an illustrative experiment comparing these two configurations and find that the histogram of collected light from a bright atom in a $\pi$-polarized trap is significantly closer to the ideal depump-free Poisson distribution than that of an atom in a $\sigma$-polarized trap \footnote{This data was collected with fiber-coupled fluorescence and is meant only to illustrate the difference between $\pi$- and $\sigma$- polarized trap light, not to be an example of detection fidelity with optimal settings.}. 
A fit to Eq.~(\ref{eq:det-model}) yields depumping probabilities per scattering event of \sigmaDepumpProbability{} and \piDepumpProbability{} for the $\sigma$ and $\pi$ configurations respectively. Uncertainties here are fitting uncertainties.
Since the other probe and trap parameters are held fixed, we infer the difference in the depump rates between the two configurations to be caused by off-resonant, V-type Raman transitions. This data was taken prior to a significant optimization of the probe beam optics (alignment and beam quality), and we attribute the remaining depump rate in the $\pi$ configuration mostly to probe imperfections. 

Achieving excellent state readout fidelity is crucial for this platform; however, from a practical standpoint, atom retention is equally important and can be a challenging problem to solve. Photon recoil during detection of a bright state atom leads to heating that can easily overcome the trap depth \cite{MeschedeDFF, WinelandLaserCooling}. To this end, we use a counterpropagating probe beam to balance the scattering force and implement an adaptive detection scheme where we apply our probe laser in a series of 5\,$\mu s$ pulses until either the threshold number of photons is collected or we reach our maximum probe duration of $250\,\mu$s \cite{ChapmanAdaptiveDet, WinelandAdaptiveDet, LucasAdaptiveDet, LeibfriedAdaptiveDet, KimAdaptiveDet, SlichterAdaptiveDet}. Since extra photons beyond threshold provide no additional information in a thresholding detection scheme, this effectively eliminates extraneous heating at no cost to the state discrimination fidelity \cite{SupplementalMaterial}. 

\begin{figure}
    \includegraphics[trim=0.5cm 0.3cm 1.5cm 0.9cm, width=0.45\textwidth]{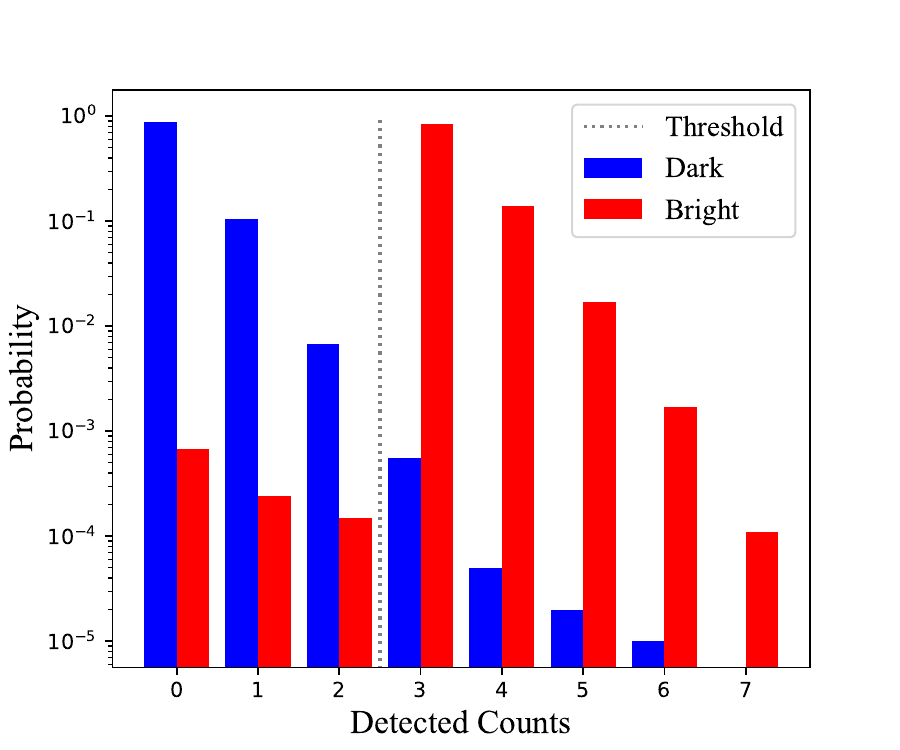}
    \caption{The normalized histogram of collected photons from the bright and dark state demonstrates \infidelity{} detection infidelity.
    Single atoms are prepared in either the dark (blue) or bright (red) state and read out by illuminating the atom with the \D{2} line probe for up to 250\,$\mu$s and collecting scattered photons. When three or more photons are collected, the atom is labelled bright; when fewer than three photons are collected, the atom is labelled dark. Atoms prepared in the bright state are erroneously labelled dark in \brightError{} of experiments, and atoms prepared in the dark state are erroneously labelled bright in \darkError{} of experiments. Data presented in this histogram was collected in 1,000 batches of 100 shots, alternating between dark and bright state preparation, for a total of 100,000 shots for each state.}
    \label{fig:det_histogram}
\end{figure}

We measure our detection fidelity by alternately preparing the atom in each state 100,000 times and recording the number of times that the measured state matches the prepared state. In Fig.~\ref{fig:det_histogram}, we show data that was collected in 1,000 batches of 100 shots, alternating between bright and dark state preparation for each batch. A total of \brightErrorN{} trials where the atom was prepared in the bright state were read out as dark, and \darkErrorN{} trials where the atom was prepared in the dark state were labelled bright. Stated as percentages, this is a bright error of $\epsilon_{\rm bright} = \brightError{}$ and a dark error of $\epsilon_{\rm dark} = \darkError{}$, yielding an infidelity of \infidelity{} or a fidelity of \fidelity{}. Uncertainties given here are Wilson score intervals. This measurement does not correct for state preparation errors, although we expect our state preparation error to be small compared to the readout error. We then infer the detection-driven loss probability by subtracting the ratio of the probability of passing the presence check when preparing in the bright and dark state from unity, and find \loss{} loss of the bright state due to detection. 

To test the applicability of our result to atom-array imaging, for which adaptive detection would be challenging using a typical camera due to slow data transfer and processing, we also measure the detection fidelity and detection-driven loss of the bright state without using adaptive detection. We reduce the probe intensity and use a probe time of 350\,$\mu s$ to achieve a bright error of $\epsilon_{\rm bright} = \freeSpaceBrightError{}$ and a dark error of $\epsilon_{\rm dark} = \freeSpaceDarkError{}$ for a combined readout infidelity of \freeSpaceInfidelity{}. We observe an additional \freeSpaceLoss{} loss of the bright state \cite{SupplementalMaterial}. We note that development and integration of single-photon avalanche diode (SPAD) array sensors could potentially provide a suitably fast signal to perform adaptive detection on an array of atoms \cite{DuttonSPAD, GersbachSPAD, GyongySPAD, RichardsonSPAD, MorimotoMegapixelSPAD}\footnote{In addition to fast imaging, individual atom addressing with single qubit rotations would be needed to optimize the enhanced survival probability from adaptive detection on an array of atoms.}.

Casting the bright state readout error as the total depump probability per collection event, we find $\mathcal{R} = 1-\sqrt[3]{1-\epsilon_{\rm bright}} = 3.7 \times 10^{-4}$. This value is the same order of magnitude as the sum of estimated contributions from off-resonant scatter from probe and trap light. The source of any remaining discrepancy in the observed depump rate is a subject of ongoing study. Possible sources of contribution to observed error rates include imperfect state preparation, overestimation of probe alignment or purity, failed presence checks resulting in poor quality post-selection, and residual $\sigma$-polarized components of the trap light due to misalignment of the trap polarization direction to the B-field or polarization distortion effects due to focusing of the trap beam. 

This depump rate is sufficiently low to permit high-fidelity detection even in the case of low collection efficiency optical systems. When we fiber-coupled our atomic fluorescence before switching to a free space detector, we still achieved \fidelityFibers{} fidelity and \lossFibers{} loss of the bright state, even though the collection efficiency was only \collectionEfficiencyFibers{}. In such a system with low depump rates and low collection efficiency, detection heating losses play a more dominant role. We found that, to some degree, we could trade atom survival probability for detection fidelity and reach a detection fidelity of \highLossFidelity{} with \highLossLoss{} loss of the bright state due to detection.

We note that further improvement of the detection fidelity and atom retention should be achievable by improving the photon collection efficiency. Our measured collection efficiency of \collectionEfficiency{} is relatively poor compared to other works \cite{SaffmanStateDet, MeschedeBayesianStateDet} and we expect that the detection infidelity should improve approximately linearly with the ratio of collected photons to depump rate. We also note that these results ease vacuum system requirements for high fidelity detection of alkali atoms in optical tweezers, since detection fidelity is independent of the background atom loss rate. To achieve comparable fidelities with a pushout method, the average atom loss rate due to background gas collisions (or indeed any source other than the pushout beam) must be comparable to the readout error reported here, since a dark atom lost due to background is indistinguishable from a bright atom lost to the pushout beam in such schemes. 
Finally, we note that our result obviates the need for toggling the trap and probe beams, as we keep the trap beam on at all times and still achieve high fidelity. Use of a magic wavelength trap is important for our ability to keep the trap on, as we avoid dipole force fluctuation (DFF) heating \cite{MeschedeDFF}. Toggling the trap and probe beam could be an alternate strategy to mitigate both the two-photon depumping mechanism and DFF heating, at the cost of potentially introducing an extra heating mechanism from repeated kicks. 

Ultimately, the fidelity reported here is the result of study of the relevant atom-photon interaction physics, and mitigation of a previously unreported depumping mechanism under simultaneous trap and probe illumination. We find that the remaining infidelity from off-resonant scattering of trap and probe light is sufficiently low to enable an order of magnitude improvement over previously published low-loss readout of alkali atoms in optical tweezers. In conjunction with the high atom retention enabled by adaptive detection, this result alludes to promise of near-term detection suitable for fault tolerant operation
and the possibility of non-disruptive mid circuit measurement for error detecting or correcting algorithms. 
This work represents an important step towards building scalable, high-performance quantum information processors and quantum sensors out of alkali-metal atoms trapped in optical tweezers.

We thank Paul Parazzoli, Ivan Deutsch, Justin Schultz, and Vikas Buchemmavari for helpful discussions.
Sandia National Laboratories is a multimission laboratory managed and operated by National Technology \& Engineering Solutions of Sandia, LLC, a wholly owned subsidiary of Honeywell International Inc., for the U.S. Department of Energy's National Nuclear Security Administration under contract DE-NA0003525.  This paper describes objective technical results and analysis. Any subjective views or opinions that might be expressed in the paper do not necessarily represent the views of the U.S. Department of Energy or the United States Government. SAND2022-10823 O 

\bibliography{main}

\pagebreak{}

\end{document}


\preprint{AIP/123-QED}

\title[]{Supplemental Material} 

\author{Matthew N. H. Chow}
\email{mnchow@sandia.gov}
\affiliation{Sandia National Laboratories, Albuquerque, New Mexico 87185, USA}
\affiliation{Department of Physics and Astronomy, University of New Mexico, Albuquerque, New Mexico 87106, USA}
\affiliation{Center for Quantum Information and Control, University of New Mexico, Albuquerque, New Mexico 87131, USA}
\author{Bethany ~J. Little}
\affiliation{Sandia National Laboratories, Albuquerque, New Mexico 87185, USA}
\author{Yuan-Yu Jau}
\affiliation{Sandia National Laboratories, Albuquerque, New Mexico 87185, USA}
\affiliation{Department of Physics and Astronomy, University of New Mexico, Albuquerque, New Mexico 87106, USA}
\affiliation{Center for Quantum Information and Control, University of New Mexico, Albuquerque, New Mexico 87131, USA}

\date{\today}

\renewcommand\figurename{FIG. S}
\makeatletter
\def\fnum@figure{\figurename\thefigure}
\makeatother

\maketitle



\section{Adaptive Detection For Loss Mitigation}
\label{section:adaptive_det}

In the main text, we emphasize techniques to mitigate various depumping mechanisms and achieve high-fidelity state detection. Here, we describe in more detail a technique to address a similarly important matter - atom retention in the trap after detection. We show that adaptive detection can reduce extraneous recoil heating, allowing for improved survival probability of bright state atoms.

Photon recoil from the near resonant probe beam causes a bright atom to undergo a random walk in momentum space, eventually allowing the atom to gain sufficient energy to overcome the trap depth ($U_{0}/k_{b}=$\trapDepthuK{}) and escape. Other heating mechanisms, such as dipole-force fluctuation heating \cite{MeschedeDFF} and ODT intensity fluctuation heating \cite{ThomasDipoleTrapHeating} are neglected here due to their small contributions on this experiment, as observed by the small AC Stark shift on the probe transition \footnote{The trap wavelength is nearly magic for the probe transition.} and long (seconds scale), trap power independent, atom storage time respectively. In a standard, single-pulse detection scheme, the mean number of counts collected from the bright state must be significantly higher than the threshold level in order to collect above-threshold counts from the bright state with high probability. This poses a problem when the recoil heating becomes comparable to the trap depth. For example, to achieve $\geq$ 99.9\% bright label probability from a Poisson count distribution by thresholding at 3 or more counts, the mean would need to be $\geq$ 11.2 counts. Dividing by our collection efficiency of $CE = \collectionEfficiency{}$ gives a requirement of $N_{scat} \approx$ 1170 photons scattered on average. If we approximate the expected heating from photon recoil with the free atom case, a single resonant scattering event adds $\Delta E/k_{b} = 2T_{\rm rec} \approx 2*0.2\,\mu$K per recoil event \cite{WinelandLaserCooling, MeschedeDFF, SteckCsNumbers}, such that the mean energy gain after single pulse detection would be 0.47\,mK. This energy gain is greater than the trap depth, implying a significant loss probability due to detection. 

However, if state labelling is determined entirely with thresholding, any extra light collected beyond the threshold level provides no additional information. Therefore, we can avoid much of the unnecessary recoil heating if we turn off the detection light as soon as we have reached the threshold level. In the ideal case of stopping the instant the threshold photon is received, the mean number of photons collected for a depump-free bright atom would simply be the threshold level, and thus the average temperature gain would be roughly a factor of $3/11.2$ times lower than the constant probe time case, keeping the average energy gain below the trap depth.

\begin{figure}
    \includegraphics[width=0.5\textwidth]{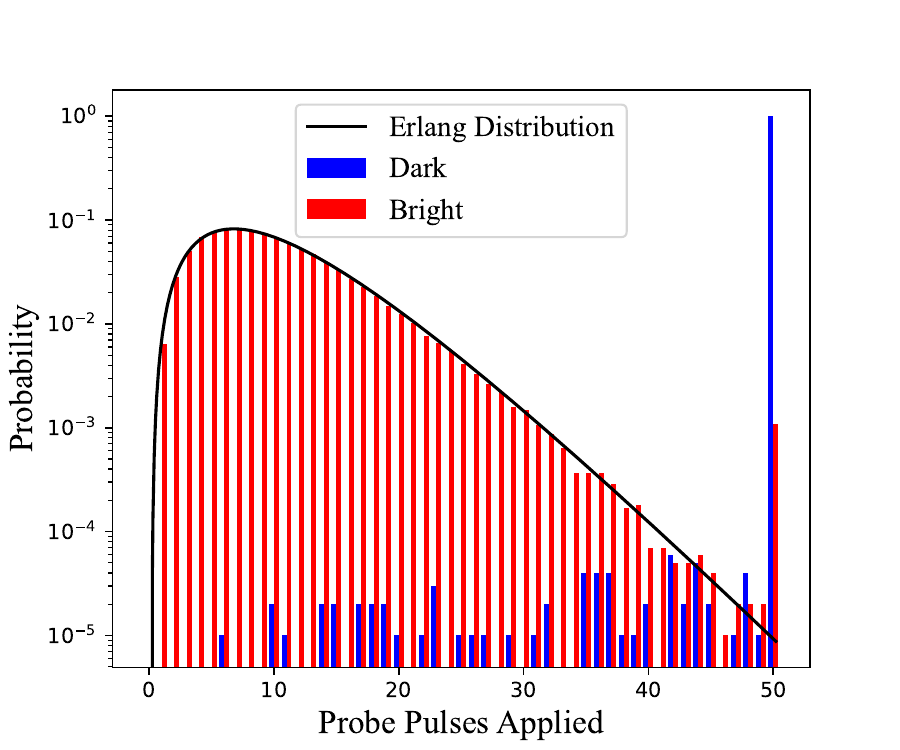}
    \caption{We measure the wait time distribution for the bright and dark state using adaptive detection. 5\,$\mu$s probe pulses are applied until the threshold level of counts are recorded or the maximum possible detection time of 250\,$\mu$s is reached. The time distribution of the bright state (red bars) closely matches an Erlang distribution (black line), which is the expected distribution for the depump-free case. The discrepancy between the chance that bright atoms that reach the maximum detection time (\brightError{}) and the Erlang distribution at max time reveals the probability that the atoms become dark before the threshold level of photons is collected. The time distribution for atoms prepared in the dark state (blue bars) is nearly unity at the maximum time bin. The combined effect of background counts on the detector and dark atoms becoming bright allows for small, but non-zero, probability of reaching threshold counts and stopping detection early when preparing in the dark state. This data was produced by recording the number of pulses applied when collecting data in Fig.~3 of the main text.}
    \label{fig:time_distribution}
\end{figure}

We refer to this protocol as ``adaptive detection,'' and note that similar protocols were used on neutral atoms in Ref. \cite{ChapmanAdaptiveDet} and on trapped ions in Refs. \cite{WinelandAdaptiveDet, LucasAdaptiveDet, LeibfriedAdaptiveDet, KimAdaptiveDet, SlichterAdaptiveDet}. We implement such a scheme by applying a series of short (5\,$\mu$s) pulses until either the threshold number (3 counts) are detected or we reach a total integrated probe duration of $t_{d} = 250\,\mu$s. 
This maps the bright state count distribution in Eq.~1 of the main text 
onto a nearly Erlang wait time ($t$) distribution with shape parameter equal to the detection threshold ($m=3$): $f(t)\approx R_{\rm bright}^{3}t^{2}e^{-R_{\rm bright}t}/2!$ shown in Fig.~S\ref{fig:time_distribution}.

\section{Probe Off-Resonant Scatter Rate} 
The calculation of the single-photon off-resonant scattering rate is done following Ref. \cite{SaffmanStateDet}. Here, the atom is assumed to start in the stretched ground state ($\ket{F=4, m_F=4}$) and we calculate the equilibrium population under probe light illumination in each excited state ($\ket{F',m'_F}$). The depump rate is then calculated by summing the scattering rate on each excited state ($R_{\rm scat,F',m'_F}$) that is reachable with a dipole-allowed transition, scaled by their branching ratios to the dark ground state manifold ($b_{F'}$) \cite{SaffmanStateDet, Metcalf}:
\begin{align}
    R_{\rm scat, F',m'_F} &= \Gamma\frac{\Omega_{F',m'_F}^{2}}{\Gamma^{2}+2\Omega_{F',m'_F}^{2}+4\delta_{F',m'_F}^{2}},
    \label{eq:rscat_two_level} \\
    R_{\rm dep,probe} &= \sum_{F',m'_F}b_{F'}R_{\rm scat, F',m'_F}~~,
    \label{eq:rdep_probe}
\end{align}
where $\Gamma$ is the natural linewidth of the $D_2$ transition, $\delta_{F',m'_F}$ is the probe laser detuning from $\ket{F',m'_F}$, and ${\Omega_{F',m'_F} \equiv \Omega_{F',m'_F,F,m_F}|_{F=4, m_F=4}}$ is the optical Rabi rate associated with the transition from state $\ket{F=4, m_F=4}$ to state $\ket{F',m'_F}$. 
We find the $D_2$ optical transition Rabi frequency between a ground-state hyperfine sublevel $|F, m_{F}\rangle$ and a excited-state hyperfine sublevel $|F', m_{F}'\rangle$ to be \cite{JauBook, SaffmanStateDet}
\begin{align}
    \begin{split}
   \Omega_{F', m_{F}',F, m_{F}}&=\sqrt{\frac{2}{3}}\frac{\mathcal{E}_{q}d}{\hbar}\times(-1)^{2F'+F-q+7}
   \sqrt{(2F+1)(2J + 1)}\\
   &\times \begin{Bmatrix} 
          J & I & F \\ 
          F' & 1 & J'
       \end{Bmatrix}_{6j}\mathcal{C}^{F',m'_{F}}_{F,m_F,1,q}~~,
    \end{split}
    \label{eq:rabi_rate}
\end{align}
where $\mathcal{E}_q$ is the spherical component of the electric field with polarization $q=m'_{F}-m_{F}$, $F=4 ~(F')$ and $J=1/2  ~(J'=3/2)$ are the ground (excited) state hyperfine and total electron angular momentum quantum numbers, $I=7/2$ is the nuclear spin, $\mathcal{C}^{F'm'_F}_{F m_F 1 q}$ is a Clebsch-Gordan coefficient, and $\{\}_{6j}$ is a Wigner-6j symbol. We use the subscripts $\{-,z,+\}$ to denote $q=\{-1,0,+1\}$. The electric dipole moment, $d$, is calculated as
\begin{equation}
   d=e\int_0^\infty P_{n'l'J'}^\ast(r)rP_{nlJ}(r)dr.
    \label{eq:dm}
\end{equation}

Here, $e$ is the unit charge, and $P_{nlJ}(r)$ is the normalized radial wavefunction. For the Cs $D_2$ transition, we have $n=6$, $l=0$, $J=1/2$ as $6S_{1/2}$ and $n'=6$, $l'=1$, $J'=3/2$ as $6P_{3/2}$ \footnote{For calculations using representation of reduced matrix elements, we find $\langle  6P_{3/2}\|er\|6S_{1/2} \rangle=\sqrt{2/3}d$ for the convention used in Ref.\cite{SteckCsNumbers,BrinkBook}, and $\langle  6P_{3/2}\|er\|6S_{1/2} \rangle=\sqrt{4/3}d$ for the convention used in Ref.\cite{JauBook,VarshalovichBook, LeKien2013}.}.

\begin{figure}
    \includegraphics[width=0.45\textwidth]{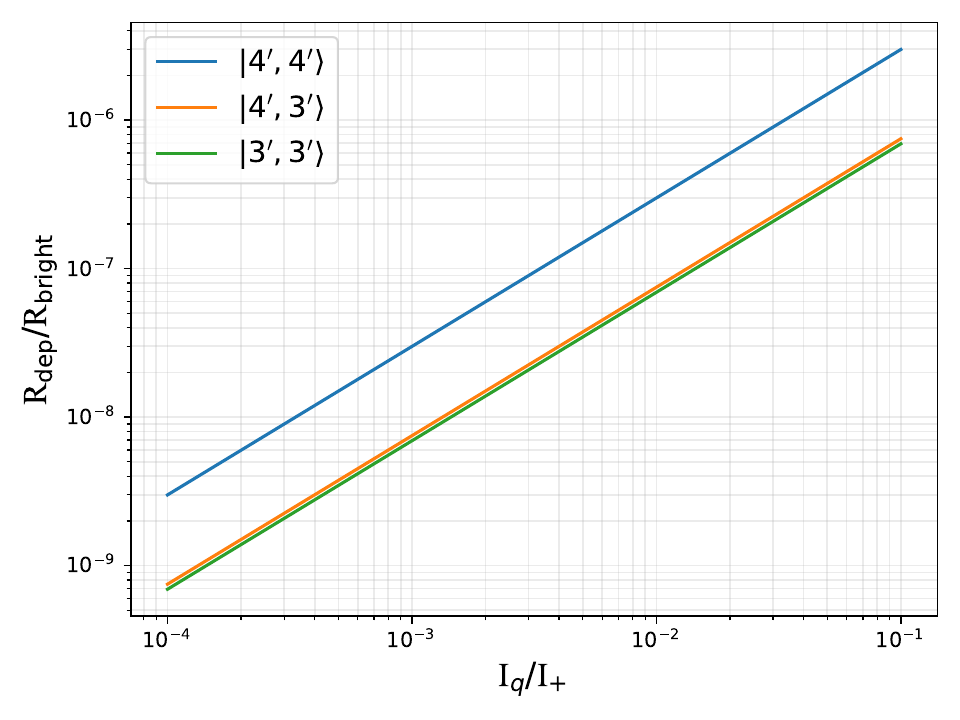}
    \includegraphics[width=0.45\textwidth]{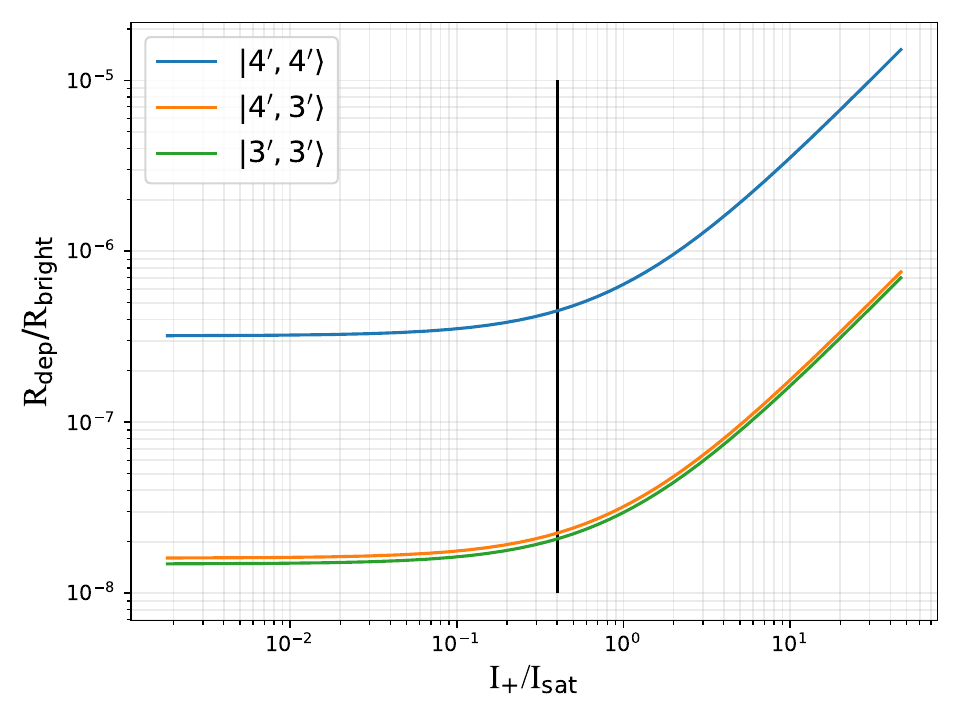}
    \caption{Numerical results for the probe off-resonant scattering rate as a function of probe polarization components (top) and probe intensity (bottom) imply the expected contribution from each transition to detection error near the range of experiment parameters. Nominal probe intensity used in the experiment of $I_{+}/I_{\rm sat} = 0.40$ is used in calculation of the top plot and is marked by a vertical line in the bottom plot. $I_{\rm sat}$ is the saturation intensity on the $\ket{F=4,m_F=4} \rightarrow$ $\ket{F'=5',m'_{F}=5'}$ transition.  Polarization fractions of $I_{z}=1.5\%$ and  and $I_{-}=0.3\%$ are used in the bottom plot.}
    \label{fig:OffRes}
\end{figure}

The Rabi rates for transitions driven by probe polarization components ($q\in\{-1, 0\}$) may then be written in terms of a polarization intensity ratio ($\Tilde{I}_{q} = |\mathcal{E}_{q}|^2/|\mathcal{E}_{+}|^2$), a state dependent coefficient ($\xi_{F',m'_F}$), and the Rabi rate on the stretched state transition ($\Omega_{5',5'}$). This expression is
\begin{equation}
    \left|\Omega_{F',m'_{F}}\right|^{2} = \Tilde{I}_{q}\xi_{F',m'_{F}}\left|\Omega_{5',5'}\right|^{2},
\end{equation}

where
\begin{align}
\begin{split}
    \xi_{F',m'_{F}} &= \frac{
    \left|\mathcal{C}^{F'm'_F}_{F m_F 1 q}
        \begin{Bmatrix} 
          J & I & F \\ 
          F' & 1 & J'
       \end{Bmatrix}_{6j}\right|^{2}}
       {\left|\mathcal{C}^{5~5}_{4~4~1~1}
        \begin{Bmatrix} 
          \frac{1}{2} & \frac{7}{2} & 4 \\ 
          5 & 1 & \frac{3}{2}
       \end{Bmatrix}_{6j}\right|^{2}} \\
       &= 36\left|\mathcal{C}^{F'm'_F}_{F m_F 1 q}
        \begin{Bmatrix} 
          J & I & F \\ 
          F' & 1 & J'
       \end{Bmatrix}_{6j}\right|^{2}.
   \label{eq:xi}
\end{split}
\end{align}
Numerical values of $\xi_{F',m'_{F}}$ are summarized in Table~\ref{Tab:off_res} for the relevant transitions. We numerically carry out the sum in Eq.~\ref{eq:rdep_probe} using probe detuning of $\delta_{5',5'} = -0.26$\,MHz from the target transition (the detuning used in the main text experiment) for a range of polarization intensity ratios and total intensities. The results from these calculations are shown in Fig.~S\ref{fig:OffRes}. 

We measure the probe light to have 99.4\% degree of polarization out of the fiber launch, (before incidence on the gold mirror) and use a polarimeter to maximize the $\sigma_{+}$ polarized component. 
Other possible contributions to polarization imperfections come from the 45$\degree$ angle of incidence, unprotected gold mirror (Thorlabs PF05-03-M03) and any potential angular misalignment between the probe propagation direction and the bias magnetic field (B-field). 
The gold mirror has a specified nominal reflectance difference of 1.4\% at 852\,nm for S- versus P-polarized light, which leads to a small $\sigma_{-}$ component. We find the magnitude of this component by taking the mirror to be a partially polarizing reflector, with Jones matrix:
\begin{equation}
    \label{eq:partial_polarizer}
    M = \begin{bmatrix}
      \sqrt{r_{s}} & 0 \\
      0 & -\sqrt{r_{p}}
    \end{bmatrix}
\end{equation}
in the s-p basis, where $r_{s}$ ($r_{p}$) is the reflectance for S-polarized (P-polarized) light. We then take the input state to be pure $\sigma_{+}$-polarized, 
calculate the renormalized output state, and project it onto the $\sigma_{-}$ basis vector. This yields an intensity fraction of $\sigma_{-}$ light relative to total intensity:
\begin{equation}
    \frac{I_{-}}{I} = \frac{r_s + r_p - 2\sqrt{r_sr_p}}{2(r_s + r_p)}.
\end{equation}
For the gold mirror's expected reflectance, $r_s = 0.98031837$, $r_p =0.96616158$, and $I_{-}/I \approx 1.3 \times 10^{-5}$.

For small misalignment angle $\theta$ between the probe propagation direction and the B-field, there is a $\sin(\theta)/\sqrt{2} \approx \theta/\sqrt{2}$ $\pi$-polarized component and $(\cos(\theta)-1)/2\approx -\theta^{2}/4$ $\sigma_{-}$-polarized component of the electric field. We conservatively estimate our mechanical angular tolerance to be  $\leq 10\degree$, contributing at most $\approx 1.5\%$ and $\approx 0.006\%$ to the $\pi$ and $\sigma_{-}$ intensity ratios respectively. Thus, $\pi$ polarization components constitute the bulk of the polarization imperfections from tilt misalignment. Evaluating Eq.~\ref{eq:rdep_probe} using $\Tilde{I}_{z}=1.5\%, ~ \Tilde{I}_{-} = 0.3\%$, and the intensity used in the result of the main paper data ($I_{+}/I_{\rm sat} = 0.40$), we predict $R_{\rm dep}/R_{\rm bright} \approx 4.9\times 10^{-7}$, or $\mathcal{R}_{\rm probe} \approx 5.1 \times 10^{-5}$ after accounting for collection efficiency.

\begin{table}
    \centering
    \begin{tabular}{ |c|c|c|c|c| } 
     \hline
     $F', m_{F}'$ & Polarization & $\delta_{F'}/2\pi$ (MHz) & $b_{F'}$ & $\xi_{F',m'_F}$ \\ 
     \hline
      4, 4 & $\pi$ & 251.00(2) & 5/12 & 7/15 \\ 
     \hline
      4, 3 & $\sigma_{-}$ & 251.00(2) & 5/12 & 7/60 \\ 
     \hline
      3, 3 & $\sigma_{-}$ & 452.24(2) & 3/4 & 7/36 \\ 
     \hline
    \end{tabular}
    \caption{\label{Tab:off_res} Summary of relevant parameters for calculation of probe-driven off-resonant coupling to the $\ket{F=4, m_{F}=4}$ ground state. $F', m'_F$ specify the excited state sublevel, $\delta_{F'}/2\pi$ is the detuning of the probe beam from that transition, $b_{F'}$ is the branching ratio to the \textit{F}=3 ground state manifold, and $\xi_{F',m'_F}$ is the angular momentum coupling coefficient normalizing the transition strength to that of the $\ket{F=4,m_F=4} \rightarrow$ $\ket{F'=5',m'_{F}=5'}$ transition, as specified in Eq.~\ref{eq:xi}. The probe is mostly $\sigma_{+}$-polarized, pumping population into $\ket{F=4, m_{F}=4}$, so only excited states with dipole-allowed coupling to $\ket{F=4, m_{F}=4}$ and non-zero $b_{F'}$ are considered. Detunings are taken from Ref.~\cite{SteckCsNumbers} assuming a resonant probe and no Zeeman splitting. Branching ratios are calculated from transition strengths given in Ref.~\cite{Metcalf} and confirmed with a formula in Ref.~\cite{SaffmanStateDet}, and $\xi_{F',m'_{F}}$ are calculated in Mathematica.}
\end{table}

\section{Trap Two-Photon Induced Depumping When a Probe is Present}
For illustration, we consider the 
scenario of the trap light induced depumping when a probe is present via the V-type stimulated Raman transitions as shown in Fig.~2c in the main body of the paper. When the bias magnetic field is orthogonal to the polarization plane of the linearly polarized trap beam, $\sigma_{+}$  and  $\sigma_{-}$ polarization components with equal electric field amplitude $\mathcal{E}_+=\mathcal{E}_-=\mathcal{E}/\sqrt{2}$ are generated by the trap light. The two-photon-transition Rabi frequency $\Omega_{\rm eff}$ caused by the $\sigma_{+}$  and  $\sigma_{-}$ transitions can be found as  \cite{Foot}
\begin{equation}\label{Oeff}
    \Omega_{\rm eff} = \frac{\Omega_{+}\Omega_{-}}{2\Delta},
\end{equation}
where $\Delta$ is the trap laser detuning from $\ket{F=4, m_{F}=4}$ level and $\Omega_{+}$ ($\Omega_{-}$) is the optical Rabi rate for each arm $\ket{F=4, m_{F}=4}$ to $\ket{F'=5, m_{F}'=5}$ ($\ket{F', m_{F}'=3}$) of the V-type Raman transition. If $\Omega_{\rm eff}$ is strong enough, the population transfer between  $\ket{F'=5, m_{F}'=5}$ and $\ket{F'=4\text{ or }3, m_{F}'=3}$ caused by the Raman transition can be non-negligible even when these hyperfine sublevels are separated by hundreds of MHz (i.e. largely RF detuned). For state detection, the probing time scale is usually much longer than the decoherence rate $\gamma_c$ of the superposition state of the excited-state sublevels. Solving the optical Bloch equation \cite{JauBook} we find the late-time ($t\gg1/\gamma_c$), mutual poluation transfer rate $\Gamma_R$ to be
\begin{equation}
    \Gamma_R=\frac{\Omega_{\rm eff}^2}{2(\delta^2+\gamma_c^2)}\gamma_c.
\end{equation}
Here, $\delta$ is the detuning that is determined by the energy difference of the associated sublevel. When the probe is present, the population ($\rho_{5',5'}$) in the source sublevel ($\ket{F'=5, m_{F}'=5}$) is nonzero, and we find the target sublevel ($\ket{F', m_{F}'}$) quasi-steady population to be  
 \begin{equation}
    \rho_{F', m_{F}'}=\frac{\Gamma_R}{\Gamma_R+\Gamma_s}\rho_{5',5'},
\end{equation}
where $\Gamma_s$ is the excited-state decay rate due to spontaneous emission. In our case, $\gamma_c=\Gamma_s$. Thus, for $\delta\gg\gamma_c$, we find 
 \begin{equation}
    \rho_{F', m_{F}'}\approx\frac{\Omega_{\rm eff}^2}{2\delta^2+\Omega_{\rm eff}^2}\rho_{5',5'}.
\end{equation}

To calculate the Raman Rabi rate $\Omega_{\rm eff}$ from a measured quantity, we find that $\Omega_{\rm eff}$ can be derived from the trap depth, $U_0$, of the far-detuned optical dipole trap, where 
\begin{equation}
   U_{0}=\frac{\hbar\Omega_{D_2}^{2}}{4} \left(\frac{1}{\Delta_{D_2}} + \frac{f_{D_1}/f_{D_2}}{\Delta_{D_1}}\right).
    \label{eq:trap_acs}
\end{equation}
Here, $\Omega_{D_2}$ is the nominal optical Rabi frequency of the linearly polarized trap light, and $f_{D_1}$ and $f_{D_2}$ are the oscillator strength for $D_1$ and $D_2$ transitions. Typically, $f_{D_1}/f_{D_2}\approx1/2$. The optical detunings $\Delta_{D_1}$ and $\Delta_{D_2}$ are defined by the center of gravity of the ground state and excited states (i.e. ignoring the hyperfine structure). Since the trap light is very far detuned, we can fairly say that $\Delta\approx\Delta_{D_2}$ for $\Delta$ in Eq.~(\ref{Oeff}).

To relate $\Omega_{D_2}$ to the electric field amplitude, we perform a more detailed calculation according to Ref.~\cite{JauBook} and find
\begin{equation}\label{OD2}
   \Omega_{D_2}=\sqrt{\frac{2}{9}}\frac{\mathcal{E}d}{\hbar},
\end{equation}
where $\mathcal{E}$ is the electric-field amplitude of the linearly polarized trap light, and $d$ is given by Eq.~\ref{eq:dm}. In our illustrative case under study, we have $\mathcal{E}_+=\mathcal{E}_-=\mathcal{E}/\sqrt{2}$. Substituting into Eq.~\ref{eq:rabi_rate}:
{\setlength{\belowdisplayskip}{0pt} \setlength{\belowdisplayshortskip}{0pt}
\setlength{\abovedisplayskip}{5pt} \setlength{\abovedisplayshortskip}{0pt}
\begin{align}\label{eq:Opm}
    \begin{split}
       \Omega_{F', m_F',F, m_{F}}&=\Omega_{D_2} \times\sqrt{\frac{3}{2}}(-1)^{2F'+F-q+7}\sqrt{(2J+1)(2F+1)}
       \\ &\times \begin{Bmatrix} z
              J & I & F \\ 
              F' & 1 & J' 
           \end{Bmatrix}_{6j}\mathcal{C}^{F',m'_{F}}_{F,m_{F},1,q}~~.
    \end{split}
\end{align}}
{\setlength{\abovedisplayskip}{0pt} \setlength{\abovedisplayshortskip}{0pt}
\setlength{\belowdisplayskip}{5pt} \setlength{\belowdisplayshortskip}{0pt}
\hspace{-0.55cm} We find
\begin{align}
    \begin{split}
        \Omega_{F', m_F'} &=\Omega_{D_2} \times\sqrt{\frac{3}{2}}(-1)^{-q+7}\sqrt{(2)(9)}
       \\ &\times \begin{Bmatrix} 
              \frac{1}{2} & \frac{7}{2} & 4 \\ 
              F' & 1 & \frac{3}{2} 
           \end{Bmatrix}_{6j}\mathcal{C}^{F',4+q}_{4,4,1,q}~~,
    \end{split}
\end{align}
for $\Omega_\pm=\Omega_{F', m_F'}$.} Definitions in Eq.~\ref{eq:Opm} are the same as in Eq.~\ref{eq:rabi_rate} and again ${\Omega_{F',m'_F} \equiv \Omega_{F',m'_F,F,m_F}|_{F=4, m_F=4}}$.

For the Raman transition from $\ket{F'=5, m_{F}'=5}$ to $\ket{F'=4, m_{F}'=3}$, we find the Rabi rate for the two arms of the V-type Raman transition to be $\Omega_{+} = \sqrt{\frac{3}{2}}\sqrt{\frac{1}{2}}\Omega_{D_2}$ and $\Omega_{-} = -\sqrt{\frac{3}{2}}\sqrt{\frac{7}{120}}\Omega_{D_2}$ from Eq.~(\ref{eq:Opm}). Therefore,
\begin{equation}
    \Omega_{{\rm eff}, 4'} =  -\sqrt{\frac{21}{320}}\frac{\Omega_{D_2}^{2}}{2\Delta_{D_{2}}},
\end{equation}
and from Eq.~(\ref{eq:trap_acs}), we find 
\begin{align}
    \label{eq:omega_from_depth}
    \Omega_{D_2}^2 = \frac{4U_{0}}{\hbar}\left|\frac{\Delta_{D_1}\Delta_{D_2}}{\Delta_{D_1}+(f_{D_1}/f_{D_2})\Delta_{D_2}}\right|,
\end{align}
such that we can express the effective Raman Rabi rate directly in terms of the trap depth:
\begin{align}
    \label{eq:omega_eff_from_depth}
    \Omega_{\rm eff,4'} &= -\sqrt{\frac{21}{320}}\frac{2U_{0}}{\hbar}\left|\frac{\Delta_{D_1}}{\Delta_{D_1}+(f_{D_1}/f_{D_2})\Delta_{D_2}}\right| \\
    &\approx 2\pi \times \left(-0.255\frac{U_{0}}{h}\right).
\end{align}
Using the trap depth $U_0/h = 8.9$\,MHz and the optical detunings of the 937\,nm trap laser, we find the population $\rho_{4',3'}$ of the $\ket{F'=4, m_{F}'=3}$ excited state based on the effective Rabi rate and detuning ($\delta_{4,5}/2\pi = 251\,$MHz \cite{SteckCsNumbers}, see the definition in Fig.~1 of the main text) 
of the Raman transition to be 
\begin{align}
    \rho_{4',3'} &\approx \rho_{5',5'}\frac{\Omega_{\rm eff}^{2}}{2\delta_{4,5}^{2} + \Omega_{\rm eff}^{2}} \approx \rho_{5',5'}\times 8.2\times 10^{-5}.
\end{align}

The branching ratio of the $F'=4$ manifold to $F=3$ ground state is 5/12 (see Table \ref{Tab:off_res}), 
so the overall probability of depumping per resonant scattering event for this transition is 
\begin{equation}
    p_{\rm depump, 4',3'} = b_{4'}\frac{\rho_{4',3'}}{\rho_{5',5'}} \approx 3.4\times10^{-5}.
\end{equation}
A similar calculation can be carried out for $\ket{F'=3, m_{F}'=3}$ on the $\sigma_{-}$ arm:
\begin{align}
    \Omega_{-} &=  \sqrt{\frac{3}{2}}\sqrt{\frac{7}{72}}\Omega_{D_2}=\sqrt{\frac{7}{48}}\Omega_{D_2} \\
    \Omega_{\rm eff} &\approx 2\pi \times \left(0.329\frac{U_{0}}{h}\right) \\ 
    \frac{\Bar{\rho}_{3',3'}}{\Bar{\rho}_{5',5'}} &\approx 4.2\times 10^{-5} \hspace{1cm} ({\rm for}~U_{0}/h = 8.9\,{\rm MHz})\\
    p_{\rm depump, 3',3'} &= b_{3'}\frac{\Bar{\rho}_{3',3'}}{\Bar{\rho}_{5',5'}} \approx 3.2 \times 10^{-5}.
\end{align}

Combining these two depump probabilities, we have a $6.6\times 10^{-5}$ 
chance of depump for each resonant scattering event with this configuration, or $\mathcal{R}_{\rm Raman} \approx 6.8 \times 10^{-3}$ 
after accounting for our collection efficiency (CE) of \collectionEfficiency{}, which is a measured value that includes transmission of our optical system and the quantum efficiency of the detector.

For the calculation illustrated here, we only consider the case of  $\sigma_{+}$- $\sigma_{-}$ V-type Raman transition. For any other configurations of the trap light polarization and the bias magnetic field, $\pi$- $\sigma_{-}$ and/or $\pi$- $\sigma_{+}$ V-type Raman transitions can also exist. Mathematical treatments can be found through a similar manner.

\subsection{Reduced Effective Trap Intensity Due to Wavepacket Spread}

We also consider the possibility that the atomic wavepacket spread is significant compared to the trap size, which could arise naturally during detection due to recoil heating. In this case, the full trap depth $U_{0}$ in the previous equations is replaced with the average light shift experienced by the atom, $\Bar{U}$. We find $\Bar{U}$ by integrating the trapping potential ($U(\Vec{r})$) over the atomic probability density function ($\rho(\Vec{r})$):
\begin{equation}
    \label{eq:ubar}
    \Bar{U} = \int U(\Vec{r}) \rho(\Vec{r})d^{3}\Vec{r}.
\end{equation}

\begin{figure}
    \centering
    \includegraphics[width=0.4\textwidth, trim=1.6cm 0cm 0cm 0cm]{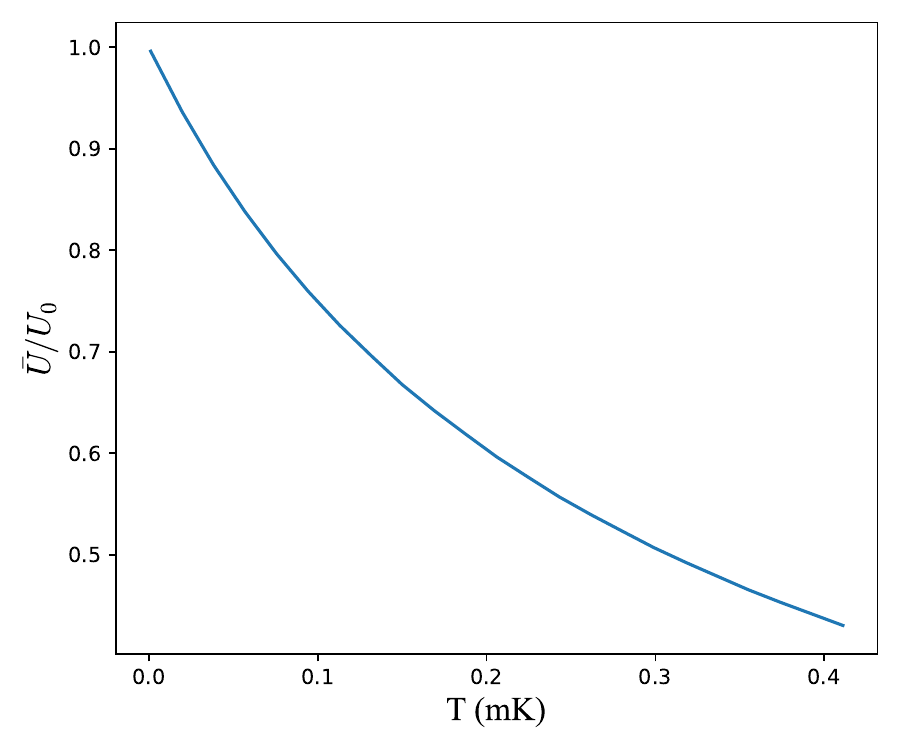}
    \caption{Average trap intensity experienced by the atom decreases due to thermal wavepacket spread as the temperature of the atom increases. We calculate $\Bar{U}/U_{0}$ for a range of temperatures up to the trap depth according to Eq.~\ref{eq:ubar_approx}. We use a trap depth of $U_0/k_b = $\trapDepthuK{}, a waist of $w_0=$1.6\,$\mu$m, and numerically integrate over $\pm 4w_0$ and $\pm 4z_R$ in the radial and axial dimensions respectively. When the atom temperature is significant compared to the trap depth, this reduction in effective intensity experienced by the atom should be accounted for when calculating both the two-photon depumping and single-photon off resonant scattering mechanisms discussed in the text.}
    \label{fig:ubar}
\end{figure}

We calculate for the case of a Gaussian beam trap with Rayleigh range $z_{R}$ and waist $w_{0}$, and we approximate the wavepacket with a thermal state for a harmonic oscillator. We note that this is only a first order approximation of the wavepacket, and it overestimates the confinement, thereby overestimating $\Bar{U}$. In this approximation, Eq.~(\ref{eq:ubar}) is a three dimensional Gaussian overlap integral \cite{FeynmanStatMech},
\begin{align}
    \begin{split} \label{eq:ubar_approx}
        \Bar{U} &= \int U_{0}\frac{\exp(-2(x_{0}^{2}+x_{1}^{2})/w^{2}(x_{2}))}{1+(x_{2}/z_{R})^{2}} \\
        & \times \prod_{i=0}^{2}
        \sqrt{\frac{m\omega_{i}~{\rm tanh}f_{i}}{\pi\hbar}}
        \exp[\frac{-m\omega_{i}}{\hbar} x_{i}^2 ~{\rm tanh}f_{i}]dx_{i},
    \end{split} \\
    & \hspace{0.5cm} f_i = \frac{\hbar \omega_{i}}{2k_{b}T} 
    \hspace{0.5cm} {\rm and} \hspace{0.5cm}
    w(x_{2}) = w_{0}^2(1+x_2^2/z^2_r),
\end{align}
where $x_{i}$ is a Cartesian position coordinate, $\omega_{i}$ is the harmonic oscillator frequency for the $i^{th}$ direction, $m$ is the mass of the atom, $k_b$ is the Boltzmann constant, and $T$ is the temperature. We carry out this integral numerically for a range of temperatures up to the trap depth to find the dependence of $\Bar{U}/U_{0}$ on the temperature. See Fig.~S\ref{fig:ubar}. These numerics indicate that this effect is significant even for temperatures as low as $0.1 \times U_{0}/k_B$. When the temperature of the atom is variable in time, to accurately include this wavepacket spread effect in calculation of either the single- or two-photon depumping mechanisms induced by trap light, one must also have knowledge of the heating rate. Then the temperature-dependent depumping rate may be found and integrated over time to find the contribution to detection infidelity.
We deem this level of precision in calculation to be out of scope for our purposes and note that a rough estimate of the wavepacket spread effect may be found graphically by estimating the area under the square of the curve shown in Fig.~S\ref{fig:ubar}. 

\section{Standard Fluorescence Detection}

\begin{figure}
    \includegraphics[width=3.325 in]{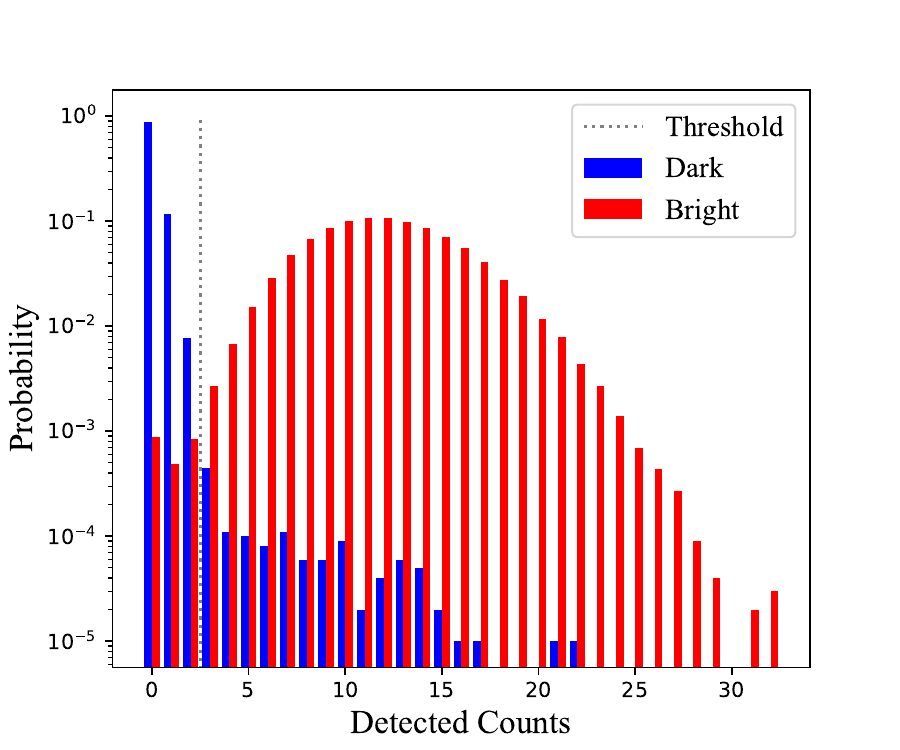}
    \caption{The normalized histogram of collected photons from the bright and dark state as measured without adaptive detection demonstrates \freeSpaceInfidelity{} detection infidelity.
    Data presented in this histogram was collected in 1,000 batches of 100 shots, alternating between dark and bright state preparation, for a total of 100,000 shots for each state.}
    \label{fig:det_histogram_fs}
\end{figure}

While adaptive detection improves the survival probability in the case of single-atom detection, direct implementation on an array of atoms would require additional technical capability. For this reason, we turn off adaptive detection and measure the detection fidelity and survival probability as an indicator for performance of our techniques on an array experiment. 
After empirically adjusting the probe power and duration for both fidelity and survival probability, we achieve a bright error of $\epsilon_{\rm bright} = \freeSpaceBrightError{}$ and a dark error of $\epsilon_{\rm dark} = \freeSpaceDarkError{}$ for a combined readout infidelity of \freeSpaceInfidelity{} in 350\,$\mu s$ probe duration. We observe an additional \freeSpaceLoss{} loss of the bright state. Full results are shown in Fig.~S\ref{fig:det_histogram_fs}. 

Similar to the case of low collection efficiency described in the main paper, in this regime survival probability may be traded for detection fidelity. In the case of optimizing only for detection fidelity, one could use the same probe settings as the main result with adaptive detection and achieve the same fidelity at the cost of significant loss of bright state atoms.

The main challenge for implementing adaptive detection on an array of atoms is the variable minimum probe duration required to determine the state of different atoms. In the case of global probe light and fluorescence collection electronics, the slowest atom determines when detection can be stopped. The presence of even one dark atom in the array would require that the entire array be exposed to probe light for the maximum duration. This challenge can be solved if two technical barriers are overcome: individual site addressing and fast site-specific photon collection signals. If each trap site can be individually added or removed from the detection process, then as soon as an atom is found to be bright, detection stops for that site. This could be done with either site-specific probe light or individual addressing single atom control (an atom found to be bright is either flipped dark with a single qubit $\pi$-pulse or shelved into a dark manifold).
In addition, fast digital processing of the fluorescence signal would be needed to avoid large overhead. Typical cameras used by contemporary tweezer-array experiments have at best millisecond timescale communication and processing time with an experiment control computer. By contrast, our SPCM communicates directly with an FPGA that does on the fly processing of incoming counts and can determine when to stop detection in a few microseconds. One could achieve similar overhead with an array of atoms by interfacing either an array of SPCMs or a single-photon avalanche diode (SPAD) array sensor directly with an FPGA \cite{DuttonSPAD, GersbachSPAD, GyongySPAD, RichardsonSPAD, MorimotoMegapixelSPAD}.  

\section{Funding Statement}

Sandia National Laboratories is a multimission laboratory managed and operated by National Technology \& Engineering Solutions of Sandia, LLC, a wholly owned subsidiary of Honeywell International Inc., for the U.S. Department of Energy's National Nuclear Security Administration under contract DE-NA0003525.  This paper describes objective technical results and analysis. Any subjective views or opinions that might be expressed in the paper do not necessarily represent the views of the U.S. Department of Energy or the United States Government. This document is SAND2022-13661 O, a supplement for SAND2022-10823 O. 

\bibliography{sm}